\documentclass[runningheads]{llncs}

\usepackage{perpage}
\usepackage{amsmath}
\usepackage{booktabs}
\usepackage{algorithm}
\usepackage{algorithmic}
\usepackage{color}
\usepackage{graphicx}
\usepackage{subfigure}

\MakePerPage{footnote}
\begin{document}
\title{LHRM: A LBS based Heterogeneous Relations Model for User Cold Start Recommendation in Online Travel Platform\thanks{Supported by National Natural Science Foundation of China (No. 61802098).}}


% \author{Ziyi Wang\inst{1} \and
% Wendong Xiao\inst{1}\orcidID{} \and
% Yu Li\inst{2}\orcidID{}}

\author{Ziyi Wang\inst{1} \and
Wendong Xiao\inst{1}\and
Yu Li\inst{2} \and
Zulong Chen\inst{1} \and
Zhi Jiang\inst{1}}


\institute{Alibaba Group, Hangzhou, China \\
\email{\{jianghu.wzy,xunxiao.xwd,zulong.cz,jz105915\}@alibaba-inc.com}\\
\and
Department of Computer Science and Technology \\Hangzhou Dianzi University, Hangzhou, China\\
\email{liyucomp@hdu.edu.cn}}
\maketitle              % typeset the header of the contribution
\begin{abstract}
Most current recommender systems used the historical behaviour data of user to predict user' preference. However, it is difficult to recommend items to new users accurately. To alleviate this problem, existing user cold start methods either apply deep learning to build a cross-domain recommender system or map user attributes into the space of user behaviour. These methods are more challenging when applied to online travel platform (e.g., Fliggy), because it is hard to find a cross-domain that user has similar behaviour with travel scenarios and the Location Based Services (LBS) information of users have not been paid sufficient attention. In this work, we propose a LBS-based Heterogeneous Relations Model (LHRM) for user cold start recommendation, which utilizes user's LBS information and behaviour information in related domains and user's behaviour information in travel  platforms (e.g., Fliggy) to construct the heterogeneous relations between users and items. Moreover, an attention-based multi-layer perceptron is applied to extract latent factors of users and items. Through this way, LHRM has better generalization performance than existing methods. Experimental results on real data from Fliggy's offline log illustrate the effectiveness of LHRM.

\end{abstract}

\section{Introduction}
Recommender Systems (RSs) aim to improving the Click-Through Rate (CTR), post-Click conVersion Rate (CVR) and stay time in the application. Most current RSs are based on the intuition that users’ interests can be inferred from their historical behaviours or other users with similar preference \cite{DBLP:conf/kdd/ZhuLZLHLG18}. Unfortunately, recommendation algorithms are generally faced with data sparsity and cold start problems so that RSs cannot guarantee high recommendation accuracies \cite{DBLP:conf/aaai/FuPWXL19,8229786}. 

Cold start problem refers to making recommendations when there are no prior interactions available for a user or an item \cite{DBLP:conf/icuimc/LamVLD08,DBLP:journals/eswa/LikaKH14,DBLP:journals/cit/Nadimi-ShahrakiB14}, which falls into two forms: (1) new user cold start problem (2) new item cold start problem \cite{8229786}. In new user cold start problem, a new user has just registered to the system and RS has no behaviour information about the user except some basic attributes \cite{SaraswathiSurvey}. In new item cold start problem, a new item is presently added to the online recommendation platform and RS has no ratings on it \cite{SaraswathiSurvey}. Compared with new item cold start problem, the new user cold start problem is more difficult and has been attracting greater interest \cite{DBLP:journals/kbs/BobadillaOHB12}. In this paper, we focus on user cold start problem. Existing methods, including cross-domain recommendation algorithms \cite{DBLP:conf/dasfaa/SongPWFHY17,DBLP:conf/dasfaa/WangPWYFH18,DBLP:conf/aaai/FuPWXL19,DBLP:conf/aaai/0004JCGCA19,DBLP:conf/www/WangAMHNYK19}, Lowrank Linear Auto-Encoder (LLAE) \cite{DBLP:conf/aaai/LiJL00H19} have been proposed and achieved great success for user cold start problem. 

User cold start recommendation over online travel platforms (e.g., Fliggy) are more challenging, thus existing methods cannot work well. LLAE \cite{DBLP:conf/aaai/LiJL00H19} can reconstruct user behavior from user attributes, but even for active user, travel is a low-frequency demand and user behaviour is quite sparse. Therefore, the generalization performance of LLAE is limited by the sparse behaviour of users. Cross-domain algorithms try to utilize explicit or implicit feedbacks from multiple auxiliary domains to improve the recommendation performance in the target domain\cite{DBLP:conf/aaai/FuPWXL19}. Unfortunately, it is hard to find a cross-domain that user has similar behaviour with travel scenarios and the LBS information of users have not been paid sufficient attention. Unconditional fuse the user behaviour information from other domains may introduce much noise. More importantly, user's travel intention is strongly related to user’ LBS information. The intuition is that, users who are geographical situation closer may have similar travel intention. 

To alleviate the user cold start problem in travel scenarios, we propose a LBS based Heterogeneous Relations Model (LHRM) for user cold start recommendation in online travel platform. LHRM firstly constructs heterogeneous relations between users and items and then apply an attention-based multi-layer perceptron to learn the latent factors of users and items. Heterogeneous relations is proposed in \cite{DBLP:journals/ipm/Cao15}, which include user-user couplings, item-item couplings, and user-item couplings. It is increasingly recognized that modeling such multiple heterogeneous relations is essential for understanding the non-IID nature and characteristics of RSs \cite{DBLP:journals/ipm/Cao15,Cao2016Non}. In order to relieve the problem of data-sparse, user behaviour information in a specific category of items in related domains is used to learn the embedding representation of user. The background is that, more than 80\% of Fliggy users have Taobao platform account\footnote{Fliggy and Taobao jointly use Taobao platform account, and relevant data sharing has been informed to users and obtained user's consent.}, and most of them are cold start users in Fliggy, but they have rich behaviours on Taobao. Then LBS information and user behaviour information in a specific category of items in Taobao domain are concatenated to construct the heterogeneous relations between users. User behaviour information in Fliggy domain is used to construct the heterogeneous relations between items. Meanwhile, user attributes are mapped into the space of user behaviour in Fliggy domain. After obtaining the side information and the embedding representation of user and items, an attention-based multi-layer perceptron is applied to extract higher level features and make the recommendation results more accurately for cold start user in Fliggy.  

In summary, the contributions of this paper are multi-fold:
\begin{itemize}
\item We propose a LBS-based Heterogeneous Relations Model (LHRM) for user cold start recommendation in travel scenarios, which utilize the LBS information and fuses user behaviour information in specific category in Taobao domain to improve the recommendation performance.
\item A new heterogeneous relations between users and items is proposed in LHRM, which can represent the relationship between users who with similar preference better.
\item Comprehensive experimental results on real data demonstrate the effectiveness of the proposed LHRM model.
\end{itemize} 

The rest of this paper is organized as follows: Section 2 review the related work. Section 3 describe the proposed model in detail. Section 4 focus on the experimental results about the proposed model, including performance evaluation on real data from Fliggy’s offline log. At last, we conclude the paper in Section 5.

\section{Related Work}
The main issue of the cold start problem is that, there is no available information can be required for making recommendations \cite{8229786}. There has been extensive research on cold start problem in recommender systems. In the section, we mainly review the related work about user cold start problem. 

Cross domain \cite{DBLP:conf/aaai/FuPWXL19,DBLP:conf/dasfaa/SongPWFHY17,DBLP:conf/aaai/0004JCGCA19,DBLP:conf/cikm/HuZY18,DBLP:conf/www/ElkahkySH15,DBLP:conf/www/WangAMHNYK19} recommendation algorithms have attracted much attention in recent years, which utilize explicit or implicit feedbacks from multiple auxiliary domains to improve the recommendation performance in the target domain. \cite{DBLP:conf/aaai/FuPWXL19} proposed a Review and Content based Deep Fusion Model (RC-DFM), which contains four major steps: vectorization of reviews and item contents, generation of latent factors, mapping of user latent factors and cross-domain recommendation. Through this way, the learned user and item latent factors can preserve more semantic information. \cite{DBLP:conf/cikm/HuZY18} proposed the collaborative cross networks (CoNet), which can  learn complex user-item interaction relationships and enable  dual knowledge transfer across domains by introducing cross connections from one base network to another and vice versa. \cite{DBLP:conf/www/WangAMHNYK19} combine an online shopping domain with information from an ad platform, and then apply deep learning to build a cross-domain recommender system based on shared users of these two domains, to alleviate the user cold start problem. 

Servel recent works model the relationship between user attributes and user behaviour. With the assumption that people with the similar preferences would have the similar consuming behavior, \cite{DBLP:conf/aaai/LiJL00H19} proposed a Zero-Shot Learning (ZSL) method for user cold start recommendation. Low-rank Linear Auto-Encoder (LLAE) consists of two parts, a low-rank encoder maps user behavior into user attributes and a symmetric decoder reconstructs user behavior from user attributes. LLAE takes the efficiency into account, so that it suits large-scale problem.

A non-personalized recommendation algorithm is proposed in \cite{DBLP:journals/is/SilvaCPMR19}. The authos hypothesize that combining distinct non-personalized RSs can be better to conquer the most first-time users than traditional ones. \cite{DBLP:journals/is/SilvaCPMR19} proposed two RSs to balance the recommendations along the profile-oriented dimensions. Max-Coverage and Category-Exploration aims to explore user coverage to diversify the items recommended and conquer more first-time users.

\section{The Proposed Approach}
\subsection{Problem Statement}
Most current Recommender Systems (RSs) based on the intuition that users’ interests can be inferred from their historical behaviours (such as purchase and click history) or other users with similar preference. However, it is difficult to recommend items to new users accurately. User cold start problem is a long-standing problem in recommender systems. In this work, we define the cold start user as the user who have not any behaviours on Fliggy in the past one month. Specifically, more than 20\% of users are cold start users in Fliggy everyday and the optimization task of user cold start is becoming very important in Fliggy. The problem can be summarized as follows:

\textbf{Problem: } Given a target domain $D_{t}$, and a source domain $D_{s}$, user $u$ is new for $D_{t}$, but it has interactions in $D_{s}$, recommend top $k$ items for $u$ in $D_{t}$.

\subsection{Notations}
In this paper, we use lowercase and uppercase letters to represent vector and matrix, respectively. We denote active users in two domain intersection as the target users $\left \{u_{t} \right \}$. For every target $u_{t}$, we denote the basic attributes as $x_{u_{t}} = \left \{ x_{1}, x_{2}, x_{3},... ,x_{k} \right \}$, the heterogeneous relation of user-user as $E_{u_{g}} = \left \{ e_{u_{1}}, e_{u_{2}}, e_{u_{3}},..., e_{u_{n}};e_{u_{t}} \right \}$, each $e_{u_{n}}$ is the representation vector of $u_{n}$. We denote the items that target users have interacted as the target items $\left \{i_{t} \right \}$. For every target $i_{t}$, we denote the basic attributes as $x_{i_{t}} = \left \{ x_{1}, x_{2}, x_{3},... ,x_{j} \right \}$, heterogeneous relation of item-item as $E_{i_{g}} = \left \{ e_{i_{1}}, e_{i_{2}}, e_{i_{3}},..., e_{i_{m}};e_{i_{t}} \right \}$, each $e_{i_{m}}$ is the representation vector of $i_{m}$. We use $U$ to denote the target user matrix, and $I$ to denote the target item matrix, $Y\in \left \{ 0,1\right \}^{\left | U \right |\ast \left | I \right |}$ be the relationship matrix between $U$ and $I$, where $Y_{u,i} = 1$ means $u_{n}$ clicked $i_{m}$.

\subsection{Geohash Algorithm}
In order to map user’ LBS information (such as latitude and longitude) to a range, we use the \textbf{Geohash} algorithm \cite{Vukovic2016HilbertGeohashH}. Geohash is a public domain geocode system invented in 2008 by Gustavo Niemeyer and G.M. Morton, which encodes a geographic location into a short string of letters and digits (e.g., geohash6(31.1932993, 121.4396019) = \textit{wtw37q}). Length of different Geohash strings represent different area of region, for example, geohash5 means a square area of about 10 square kilometers. 

\subsection{LBS based Heterogeneous Relations Model}
It is worth noting that users' interests can be inferred from historical behaviours or other users with similar preference and benefited from heterogeneous relations. Moreover, user's travel intention is strongly related to user’ LBS information (such as latitude and longitude), which based on the intuition that users who are geographical situation closer may have similar travel intention. To active this, we propose LBS based Heterogeneous Relations Model (LHRM), in which LBS information is used to construct the heterogeneous relation between users. The framework of LHRM illustrated in Fig.1. LHRM contains two modules: heterogeneous relations construction module and representation learning module. 

\begin{figure}[htbp]
	\centering
	\includegraphics[width=0.7\textwidth]{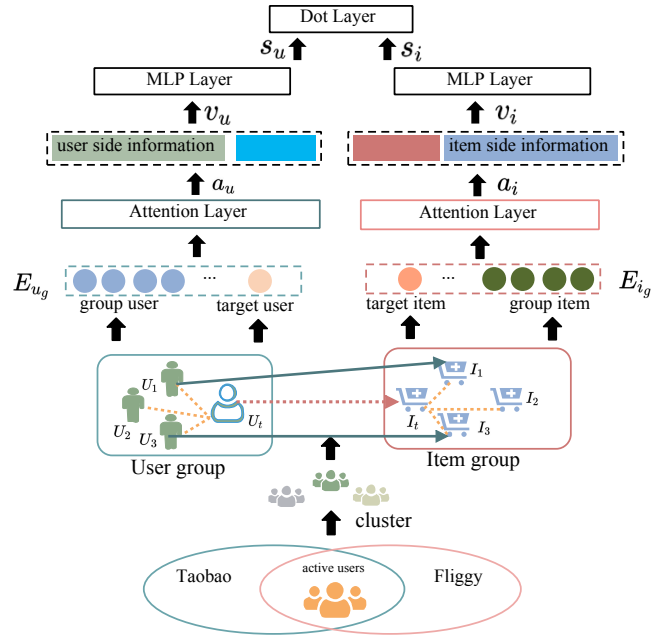}
	\caption{Framework of the proposed LHRM}\label{fig:RS-framework}
\end{figure}

\subsubsection{Heterogeneous relations construction module}
The detailed process of constructing the heterogeneous relations between users and items is shown in Fig.2. We can see that user's historical behaviours sequence and LBS sequence in Taobao domain are concatenated and input into a embedding layer, which pre-trained by skip-gram algorithm \cite{DBLP:journals/corr/abs-1301-3781}. Specifically, items not related to travel are filtered out and user's latitude and longitude information is mapped to a string with the length of 5 by the geohash5 algorithm. After the embedding layer, we adopt average-pooling to generate the corresponding vector representation of user. In order to generate different user groups, we utilize K-means algorithm to cluster users according to their representation vectors. For each user group, any user can be regarded as the target user, and the other users are regarded as the friends of the target user.

\begin{figure}[htbp]
	\centering
	\includegraphics[width=1\textwidth]{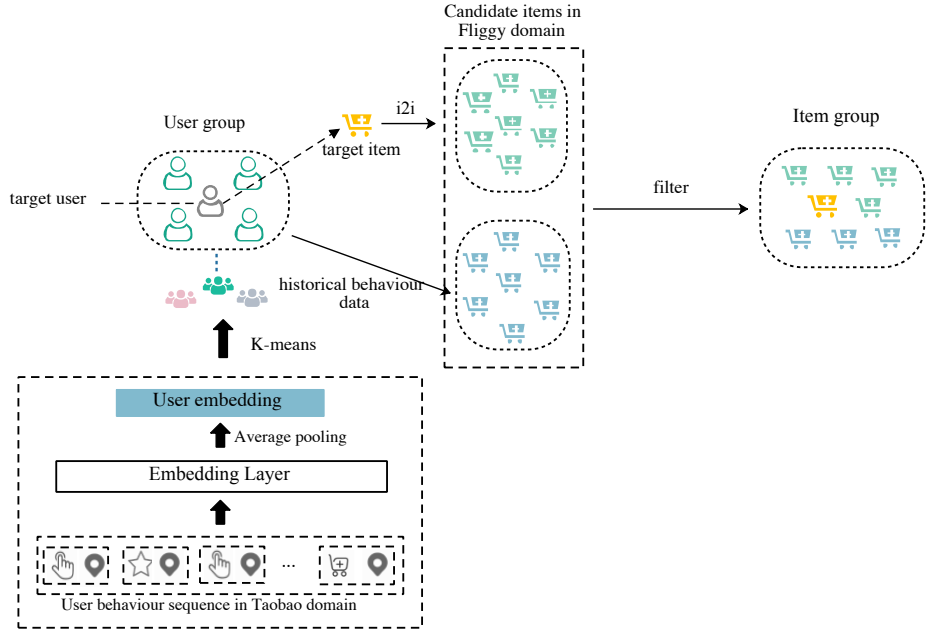}
	\caption{Architecture overview of the process to construct heterogeneous relations between users and items}\label{fig:user-group}
\end{figure}

Each item that the target user has interacted in Fliggy domain is regarded as the target item. The whole candidate items set contains two part: items recalled by target item through item-item (i2i) and items interacted by all users in user group. Finally, the items in candidate set are filtered according to the topic of the target item and generate the item group. In this way, all items in item group are more related, and which can be represented by a pre-trained item embedding vector in Fliggy domain.

\subsubsection{Representation learning module} An attention-based multi-layer perceptron is used to learn the latent factors of users and items. After the process of heterogeneous relations construction, $E_{u_{g}}$ and $E_{i_{g}}$ are generated.  Then we adopt an attention layer to focus on the relevant parts of $E_{u_{g}}$ and $E_{i_{g}}$. The implementation of attention for sequence-to-one networks on $E_{u_{g}}$ is shown in Eq.(1) and Eq.(2):

\begin{equation}
\begin{split}
\alpha _{ti}=\frac{exp(score(e_{u_{t}},e_{u_{i}}))}{\sum_{i^{'}=1}^{n+1}exp(score(e_{u_{t}},e_{u_{i^{'}}}))}
\end{split}
\end{equation}
where:
\begin{equation}
\begin{split}
score(e_{u_{t}},e_{u_{i^{'}}})=e_{u_{t}}\mathbf{W}_{a}e_{u_{i^{'}}}
\end{split}
\end{equation}
$\mathbf{W}_{a}$ is the learnable parameters in attention layer, and the output of attention layer is computed as Eq.(3):
\begin{equation}
\begin{split}
\mathbf{a}_{u}=\sum_{i=1}^{n+1}\alpha _{ti}\times e_{u_{i}}
\end{split}
\end{equation}
Similarly, attention layer is implemented on $E_{i_{g}}$, and the output is $a_{i}$. The MultiLayer Perceptron (MLP) layer is a feed-forward neural network, which can generalize better to unseen feature combinations through low-dimensional dense embeddings learned for the sparse features \cite{DBLP:conf/recsys/Cheng0HSCAACCIA16}. We denote the input of MLP layer as $v_{u}$ and $v_{i}$, the output of MLP layer as $s_{u}$ and
$s_{i}$. $v_{u} = [a_{u}, x_{{u}_{t}}]$, $v_{i} = [a_{i}, x_{{i}_{t}}]$. The output of LHRM is the preference score of $u_{t}$ for $i_{t}$, we denote $\hat{y}$ as the output of dot layer. $\hat{y}$ is computed as Eq.(4):
\begin{equation}
\begin{split}
\widehat{y}=\frac{1}{1+exp(-s_{u}\cdot s_{i})}
\end{split}
\end{equation}

$y$ is binary labels with $y=1$ or $y=0$ indicating whether click or not. The logistic loss of LHRM is shown in Eq.(5):
\begin{equation}
\begin{split}
L(y,\widehat{y})=-ylog(\widehat{y})-(1-y)log(1-\widehat{y})
\end{split}
\end{equation}

For clarity, we show the key steps of our algorithm in Algorithm 1.
\begin{algorithm}  
\caption{LBS based Heterogeneous Relations Model}  
\label{alg:A} 
\hspace*{0.02in}{\bf Input:}
User's behaviour sequence in Taobao domain $S_{T}$, user's LBS sequence in Taobao domain $L_{T}$, user's behaviour sequence in Fliggy domain $S_{F}$, user attributes $X_{u}$, item attributes $X_{i}$\\ 
\hspace*{0.02in}{\bf Output:}
Latent factors of users $S_{u}$, latent factors of items $S_{i}$
\\
\begin{algorithmic}[1]  
\STATE {$S_{T}$ and $L_{T}$ are used to construct heterogeneous relation between users by K-means, and output $E_{u_{g}}$}
\STATE {$S_{F}$ is used to construct heterogeneous relation between items, and output $E_{i_{g}}$}      
\STATE {$E_{u_{g}}$ and $E_{i_{g}}$ are input into attention layer, and output $a_{u}$, $a_{i}$}
\STATE {the concatenation vector of $[a_{u}, x_{{u}_{t}}]$ and $[a_{i}, x_{{i}_{t}}]$ are input into MLP layer, and output $s_{u}$, $s_{i}$}
\STATE {$s_{u}$, $s_{i}$ are input into dot layer, and output $\hat{y}$}
\STATE update all parameters according to $L(y,\widehat{y})=-ylog(\widehat{y})-(1-y)log(1-\widehat{y})$

\STATE{\bf Cold start:}
\STATE{$\hat{y}_{new} = s_{u_{new}} \cdot s_{i}$}
\STATE{\bf Recommendation:}
\STATE{Computing the similarity score between the new user and all candidate items, and recommend the top-k items}
\end{algorithmic}  
\end{algorithm}  

\section{Experiment}
In this section, we conduct experiments on Fliggy and Taobao's offline log dataset to evaluate the performance of LHRM and some baseline models.

\subsection{Compared Methods}
In the experiments, we compare the following methods.
\begin{itemize}
\item \textbf{Hot}: Hot is a non-personalized recommendation algorithm, which recommends items to new users according to the popularity score of item in Fliggy domain.  
\item \textbf{HERS}: Heterogeneous relations-Embedded Recommender System (HERS) is proposed in \cite{DBLP:conf/aaai/0004JCGCA19}, which based on ICAUs to model and interpret the underlying motivation of user-item interactions by considering user-user and item-item influences and can handle the cold start problem effectively.
\item \textbf{MaxCov}: Max-Coverage (MaxCov) \cite{DBLP:journals/is/SilvaCPMR19} is a non-personalized recommendation algorithm, which aims to explore user coverage to diversify the items recommended and conquer more first-time users.
\item \textbf{LHRM}: Lbs based heterogeneous relations model proposed in this paper.
\end{itemize}
Popularity score of items is very important factor in cold start recommendation, therefore, when implementing LHRM and HERS in experiments, we fuse the popularity score of items with the score of LHRM and HERS's output. Then, final preference score of $u_{t}$ for $i_{t}$ calculated in Eq.(6):
\begin{equation}
\begin{split}
\widehat{y}=Score_{LHRM/HERS} \times Pop\_Score_{item}
\end{split}
\end{equation}

\subsection{Implementation Details}
We set the maximum length of user group and item group to 10. The number of cluster center is set to 1000. The dimension of latent factors of user and item is a hyper parameter, we set it to 32, 64, 128 and 256 in the experiments. In order to evaluate the performance of proposed methods, we adopt two evaluation metrics, i.e., Hit Rate (HR@30,@50,@100,@200), Normalized Discounted Cumulative Gain (NDCG@30,@50,@100,@200). HR is a metric of shotting accurately at target items. NDCG is a cumulative measure of ranking quality, which is more sensitive to the relevance of higher ranked items.

\subsection{Datasets}
During our survey, no public datasets for user cold start recommendation in travel scenarios. To evaluate the proposed approach, we collect the offline log data from Fliggy and Taobao domain in the past one month as the dataset. Generally, impression and click sample as positive samples, impression but not click samples as negative samples. All the data have been de-identified, and the algorithm training and application related to cold start do not involve the identification of users, so as to fully protect the privacy of users. The statics of dataset is illustrated in table 1.

\begin{table}[htbp]
  \caption{Statistics of dataset. (pos - positive, neg - negative, M - Million)}\label{Table1}
  \centering
  \setlength{\tabcolsep}{3mm}{
  \begin{tabular}{l|ccc}
    \hline
          & Training & Validation & Testing \\
    \hline
    \# of samples& 7.68M& 1.56M & 1.34M\\
    \# of pos samples & 3.64M& 0.35M&0.062M \\
    \# of neg samples& 4.04M& 1.21M& 1.28M\\
    \# of users& 1.6M& 0.437M& 0.012M\\
    \# of items& 0.15M& 0.086M& 0.2M\\
    \hline
  \end{tabular}}
\end{table}

\subsection{Results}
We show the experimental results of different models in Table 2. Among all methods, LHRM achieves the best performance in terms of all metrics. Specifically, when the dimension of latent factors of user and item is set to 32, the HR and NDCG are the highest.

\begin{table}[htbp]
  \caption{Comparison of different models on dataset.}\label{Table2}
  \centering
  \setlength{\tabcolsep}{1.5mm}{
  \begin{tabular}{l|cccc|cccc}
  \hline
  	 &\multicolumn{4}{c|}{HR}&\multicolumn{4}{c}{NDCG}\\
  \hline
  	 &@30&@50&@100&@200&@30&@50&@100&@200\\
    \hline
    Hot& 0.034& 0.065 & 0.128&0.169&0.008 & 0.014&0.023 &0.029 \\
    HERS& 0.039& 0.088&	0.142&0.245& 0.011& 0.02& 0.028&0.041 \\
    MaxCov& 0.035& 0.075& 0.0993&0.1989& 0.002&0.007 & 0.024&0.036\\
    LHRM-32& \textbf{0.0754}& \textbf{0.109}& \textbf{0.184}&\textbf{0.266}&\textbf{0.022} &\textbf{0.028} &\textbf{0.039} &\textbf{0.05}\\
    LHRM-64& 0.056& 0.097& 	0.152&0.254& 0.016&0.023 &0.031 & 0.044\\
    LHRM-128& 0.0728& 0.101& 0.149&0.255&0.02 &0.025 &0.032 &0.044 \\
    LHRM-256& 0.052& 0.089& 0.15&0.253&0.014 &0.02 &0.03 & 0.043\\
    \hline
  \end{tabular}}
\end{table}

Table 2 shows the HR and NDCG on target items, and all existing methods did not work well on cold start users. Generally, in practical applications, we not only care whether the recommended items will be clicked, but also whether the recommended items are related to the target items. Therefore, we evaluate the hit rate of different models under different degree of relevance with target items. Experimental results are shown in Fig.3. We can see that, LHRM-32 is very competitive, MaxCov performs best when calculating hit rate according to whether the destination same as the target items.

\begin{figure}[H]
\centering  %图片全局居中
\subfigure[Same destination and category]{
\label{Fig.sub.1}
\includegraphics[width=0.31\textwidth]{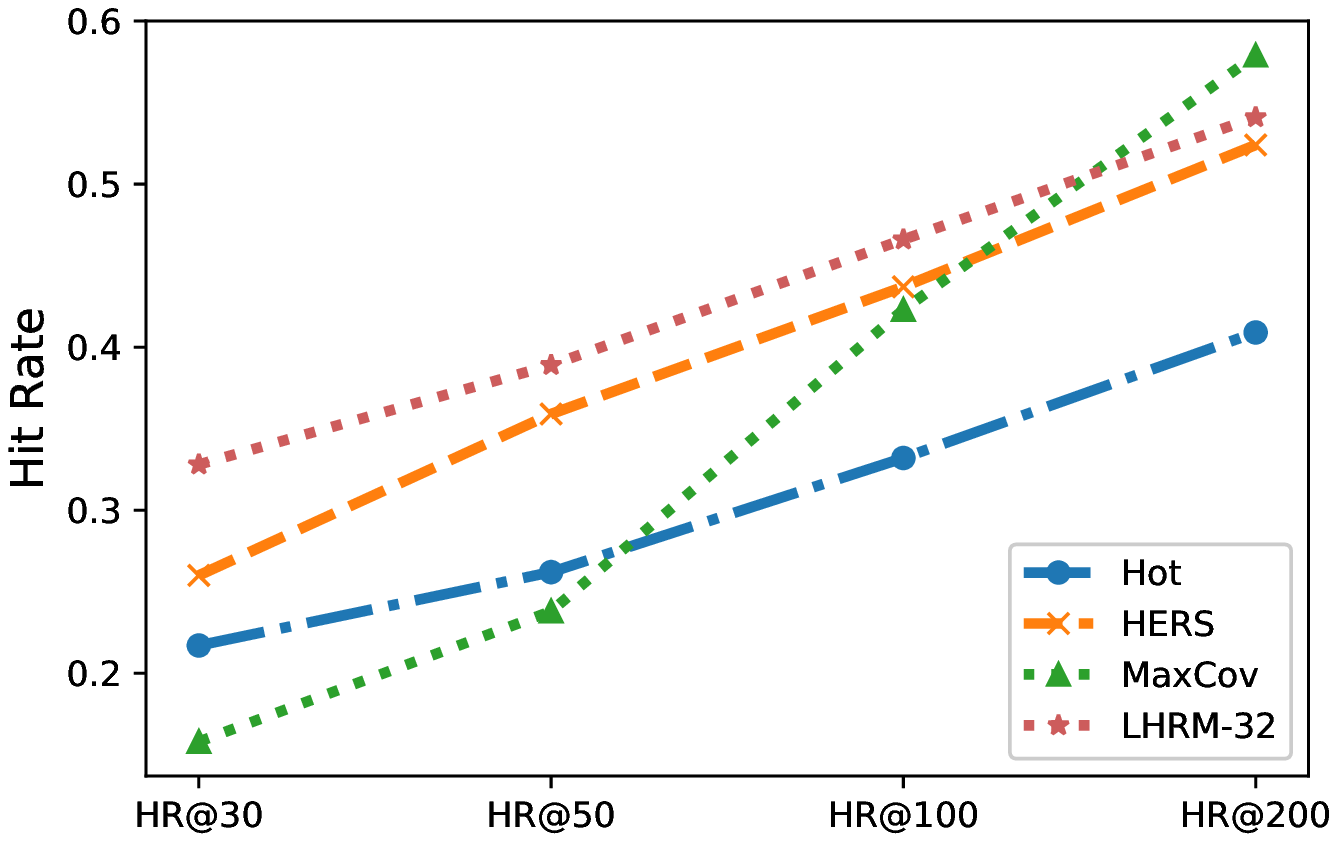}}
\subfigure[Same destination]{
\label{Fig.sub.2}
\includegraphics[width=0.31\textwidth]{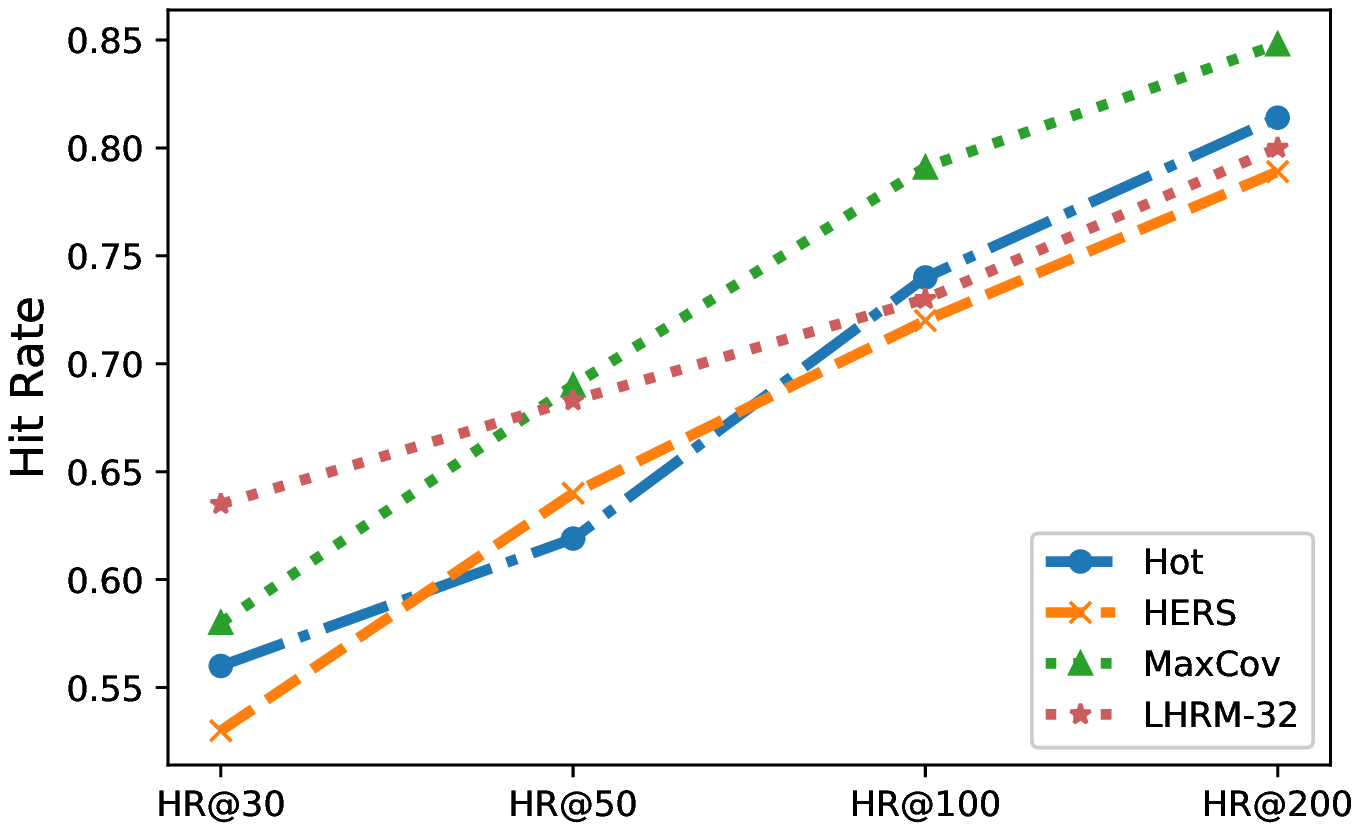}}
\subfigure[Same category]{
\label{Fig.sub.3}
\includegraphics[width=0.31\textwidth]{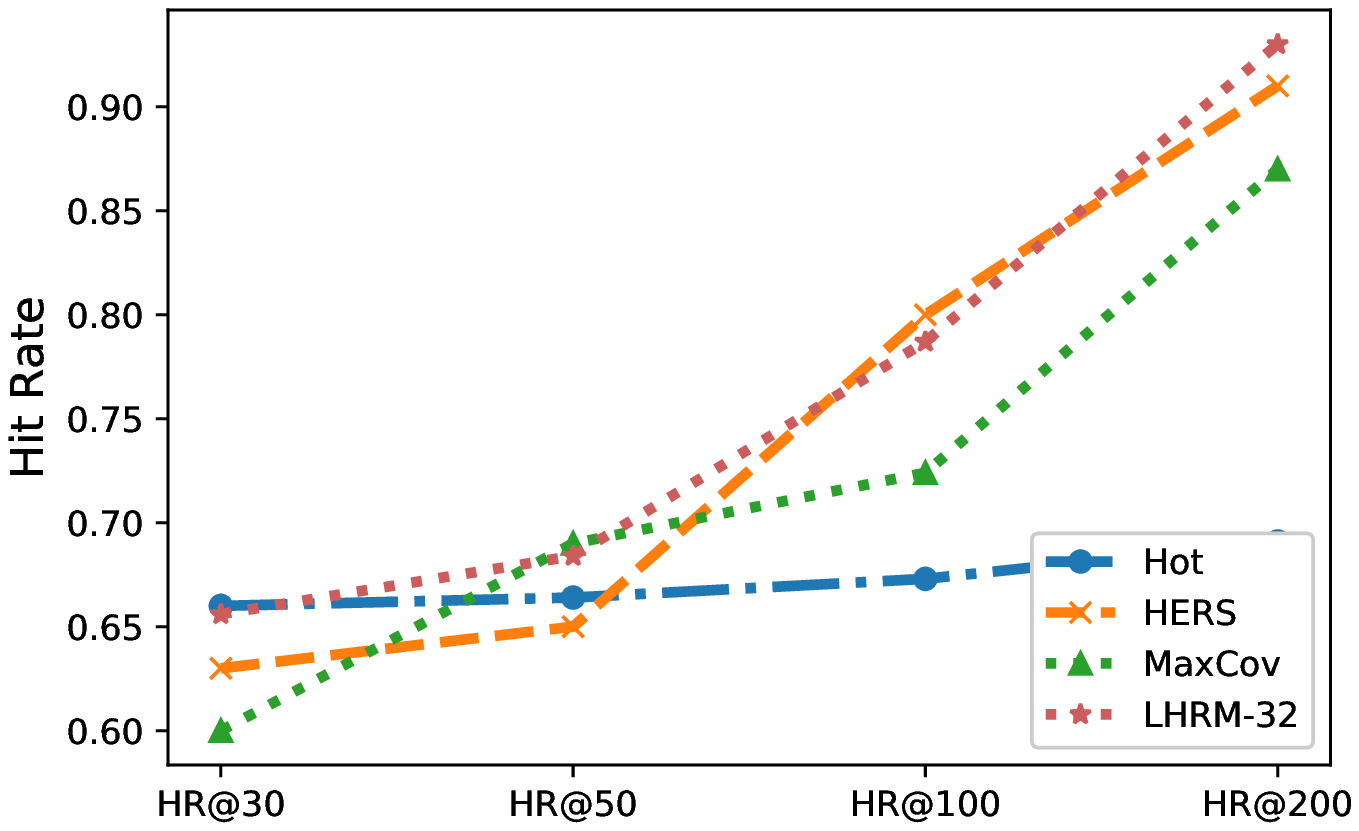}}
\caption{Comparison of different models w.r.t different degree of relevance with target items}
\label{fig:HR}
\end{figure}

\section{Conclusion}
In this paper, we point out two challenges of user cold start recommendation in travel platform: i) it is hard to find a cross-domain that user has similar behaviour with travel scenarios  ii) LBS information of users have not been paid sufficient attention. To address this problem, we propose LBS based heterogeneous relations model. LHRM utilizes user’s LBS information and behaviour information in Taobao domain and user’s behaviour information in Fliggy domain to construct the heterogeneous relations between users and items. Moreover, an attention-based multi-layer perceptron is applied to extract latent factors of users and items. Experimental results on real data from Fliggy's offline log illustrate the effectiveness of LHRM.

\bibliographystyle{named}
\bibliography{iconip20}

\end{document}